\documentclass{PoS}

\def\farcs{\hbox{$.\!\!^{\prime\prime}$}}             

\title{Spectral analyses of three carbon-enhanced metal-poor stars}

\ShortTitle{Spectral analyses of three carbon-enhanced metal-poor stars}

\author{\speaker{N.~T.~Behara$^{ab}$}, P.~Bonifacio$^{abc}$, H.-G.~Ludwig$^{ab}$, L.~Sbordone$^{ab}$,
  \hbox{J.~I. Gonz\'alez Hern\'andez$^{ab}$} and E.~Caffau$^b$\\
\llap{$^a$}CIFIST Marie Curie Excellence Team\\
\llap{$^b$}GEPI, Observatoire de Paris, CNRS, Universit\'e Paris Diderot, France\\
\llap{$^c$}Istituto Nazionale di Astrofisica,
Osservatorio Astronomico di Trieste, Italy\\
        E-mail: \email{natalie.behara@obspm.fr}}

\abstract{
Approximately 20\% of very metal-poor stars ([Fe/H] $<$ --2.0) are strongly enhanced in
carbon ([C/Fe] $>$ +1.0). Such stars are referred to as carbon-enhanced
metal-poor (CEMP) stars. A variety of abundance patterns are found among CEMP stars. Strong
overabundances of nitrogen are common, and overabundances of neutron capture
elements are often, however not always, present. The variety of abundance
patterns among CEMP stars strongly suggests that this population of stars
comprises several astrophysical origins. 

We are conducting a high-resolution follow-up
of candidate EMP stars extracted from the
Sloan Digital Sky Survey (SDSS; York et al.~2000)
using UVES at the VLT.
Three of the programme  stars,
SDSS J0912+0216, 
SDSS J1036+1212 and
SDSS J1349-0229,
where deliberately targetted as CEMP stars
since a strong $G$ band was evident from the SDSS spectra
and the weakness of the Ca\,{\sc ii} K line testified their
very low metallicity.
The UVES high resolution follow-up confirmed the original 
findings ([Fe/H] $<-2.50$ ) and allowed a more detailed investigation 
of their chemical composition.

We determined the carbon abundance from molecular lines which form
in the outer layers of the stellar atmosphere. It is known that convection in
metal-poor stars induces very low temperatures which are not predicted by
classical 1D stellar atmospheres. To obtain the correct temperature
structure, one needs full 3D hydrodynamical models. 3D carbon abundances were
determined for all three stars, using CO$^5$BOLD 3D hydrodynamical model
atmospheres. 3D effects on the carbon abundance are found to be quite significant for
these stars, with 3D corrections of up to --0.7 dex.

Two of the stars, 
SDSS J0912+0216 and 
SDSS J1349-0229
exhibit an overabundance of neutron capture elements
which classifies them as CEMP-s. Star SDSS J1036+1212,
instead belongs to the elusive class of CEMP-no/s stars,
with enhanced Ba, but deficient Sr, of which it is the third
member discovered to date.
}

\FullConference{10th Symposium on Nuclei in the Cosmos\\
		 July 27 - August 1, 2008\\
		 Mackinac Island, Michigan, USA}

\begin{document}

\section{Introduction}

\begin{figure}[ht]
\centering \includegraphics{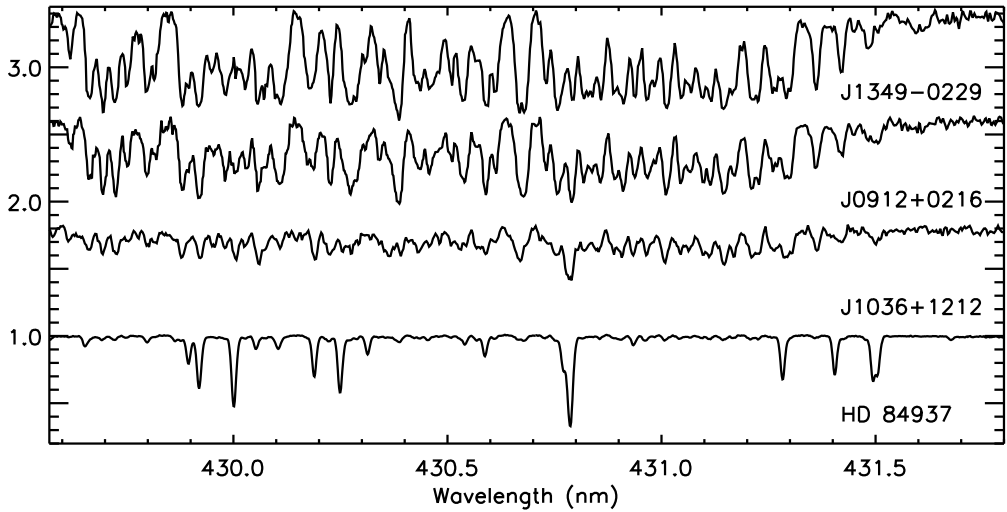}
\caption{Observed spectra of the CH {\it G} band in the three CEMP stars,
  compared to the same spectral region of a dwarf metal-poor star, HD84937,
  showing no carbon enhancement. Spectra have been offset for clarity.} 
\label{fig1}
\end{figure}

The EMP candidates have been selected using a version of our automatic analysis code 
(Bonifacio \& Caffau 2003) tailored
to SDSS spectra of turn-off stars which provides an estimate of [M/H].
Follow up high resolution spectra were acquired with UVES at the VLT,
with a 1\farcs{4} slit and $2\times2$ on-chip binning, providing
a resolution of $\sim 30\,000$.
Figure~\ref{fig1} displays the CH {\it G} band 
of all three stars, showing clearly the C enhancement.

\section{Atmospheric parameters and analysis}

As a first estimate of the effective temperatures of our stars, we used the
wings of H$\alpha$. Temperatures were then redetermined using the Fe {\sc i}
excitation equilibrium, and were found to agree with the H$\alpha$
temperatures to within 100~K. The surface gravity
was derived from the Fe {\sc i}/Fe  {\sc ii} ionisation equilibrium, and the
microturbulence was determined by requiring Fe {\sc i} lines to provide the
same abundance regardless of line strength. Equivalent widths of Fe {\sc i}
and Fe {\sc ii} lines have been measured by means of the FITLINE code
(Fran\c{c}ois et al.~2003). ATLAS model atmospheres and SYNTHE (Kurucz 1993, 2005)
synthetic spectra, in their 
Linux version (Sbordone et al.~2004, Sbordone 2005), 
have been employed in the analysis. Adopted stellar
parameters are listed in Table~\ref{tab1}. 

\begin{table}[ht]
\centering \begin{tabular}[t]{lrlrrrrr}
Star       & $T_{\rm eff}$ & log $g$  &  $\xi$  & [Fe/H] & [C/Fe]  &  [Ba/Fe] & [Sr/Fe] \\
\hline
J1349-0229 & 6200        & 4.00     & 1.5     & --3.0  & +2.17 $\pm$ 0.04  & 2.26 $\pm$ 0.05 & +1.35 $\pm$ 0.06 \\
J0912+0216 & 6500        & 4.50     & 1.5     & --2.5  & +1.53 $\pm$ 0.03  & 1.58 $\pm$ 0.04 & +0.53 $\pm$ 0.07 \\
J1036+1212 & 6000        & 4.00     & 1.4     & --3.2  & +0.98 $\pm$ 0.13  & 1.26 $\pm$ 0.03 & --0.51 $\pm$ 0.04 \\
\end{tabular}
\caption{Adopted stellar parameters and derived abundances. Carbon abundances
  were corrected for 3D effects.}
\label{tab1}
\end{table}

\section{Carbon}

\begin{figure}[ht]
\includegraphics[width=.5\textwidth]{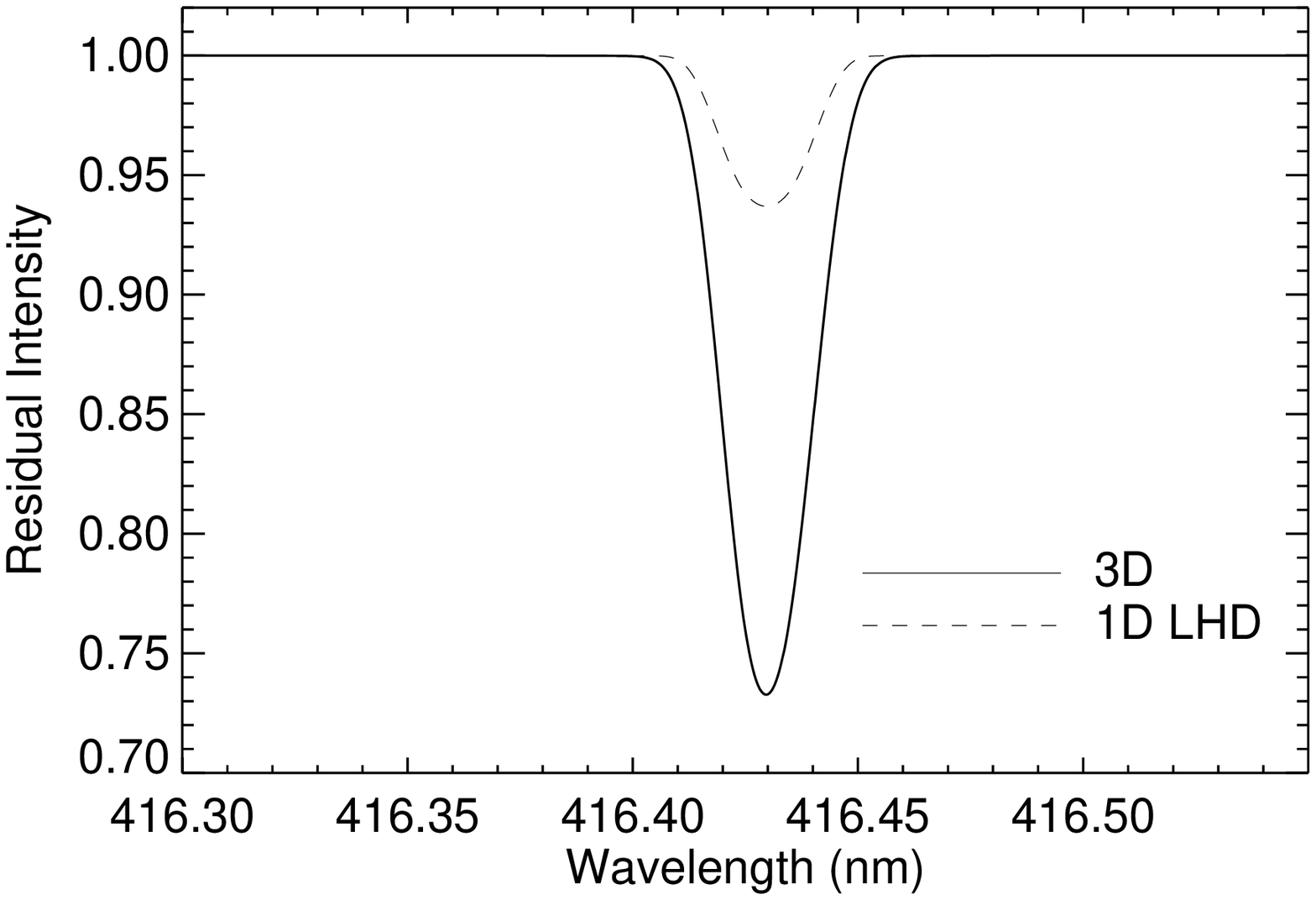}
\includegraphics[width=.5\textwidth]{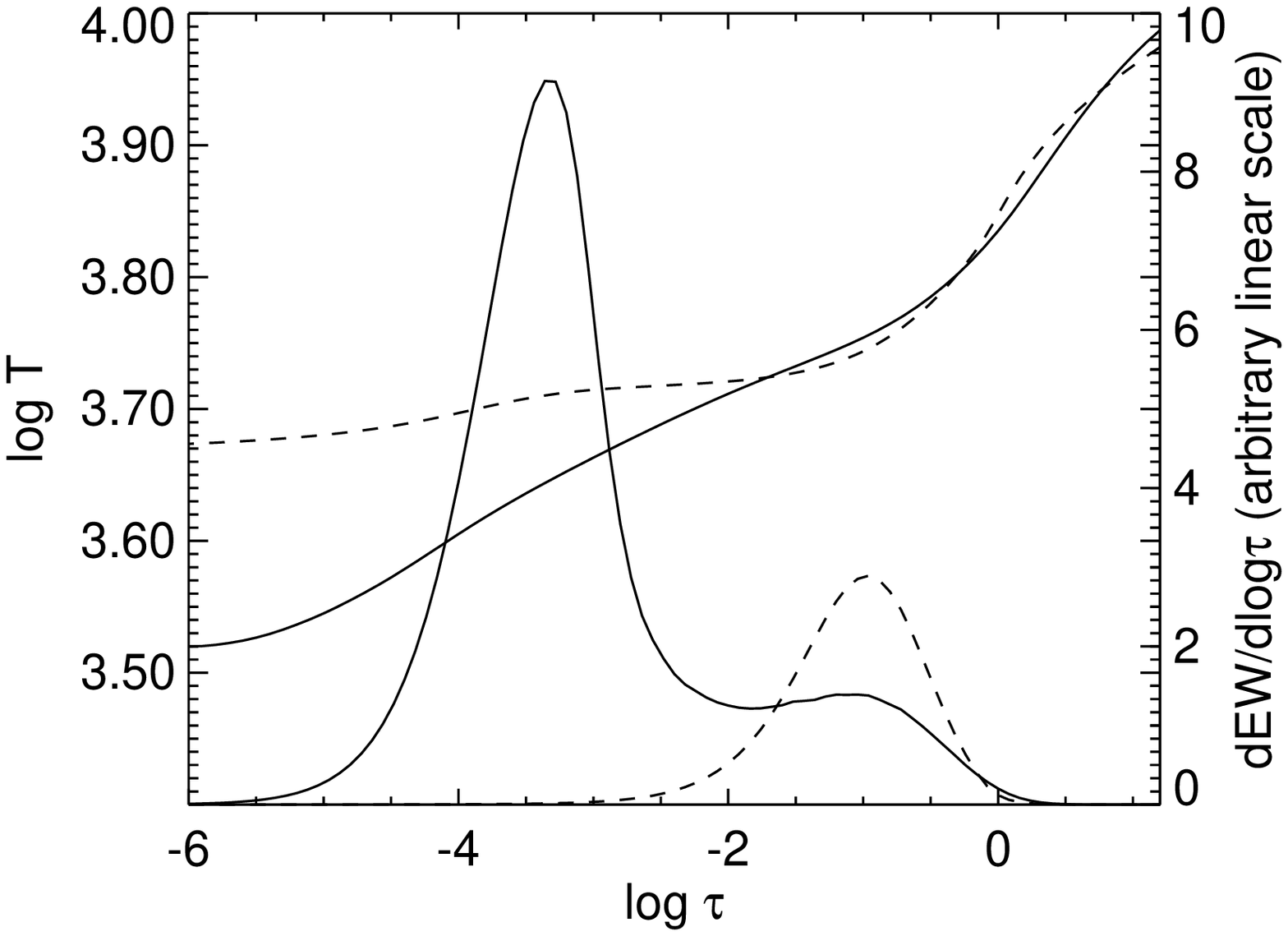}
\caption{{\it Left panel:} 3D and 1D profiles of the CH line 416.4
  nm computed using models with T$_{\rm eff}$~=~6550~K, log $g$ = 4.50 and [Fe/H] =
  --3.00. {\it Right panel:} The contribution functions for the equivalent
  width of this line, along 
  with the model temperature distributions. In both panels, the 3D model is
  plotted using a solid line and the 1D model using a dashed line.}
\label{fig2}
\end{figure}

The C abundances derived from the 1D analysis were corrected
for 3D effects (see Caffau et al.~2008, and references
therein, for our definition of 3D corrections)
using 3D model atmospheres computed with CO$^5$BOLD (Freytag et
al.~2002, Wedemeyer et al.~2004).
The 3D spectral synthesis calculations were performed with the code
Linfor3D. 
The isolated CH features at 416.4 nm, 416.9 nm, 418.0
nm, and 418.8 nm were used to determine C abundances. 
These lines form in the outer layers of the stellar atmosphere.
3D and 1D profiles of the CH line 416.4 nm are shown in the left panel of
Figure~\ref{fig2}. The contribution functions for this line, along  
with the model temperature distributions are shown in the right panel. The
outer layers of the 3D atmosphere are much cooler compared to the 1D model,
resulting in a much stronger line.
Throughout the analysis, the oxygen abundance was fixed at [O/Fe] = 0.4.

There are two main effects that distinguish 3D from 1D models,
the average temperature profile and the horizontal temperature
fluctuations. We quantify the contribution of both effects by introducing
the 3D correction as: 3D -- 1D. 
The 3D corrections obtained for the three stars range from --0.5 to --0.7 dex,
and have been already applied to the carbon abundances listed in Table~\ref{tab1}. 

\section{Neutron-capture elements}

We derive strontium abundances for our stars using the Sr {\sc ii} 407.7 nm and
421.5 nm lines. The two stars with the highest carbon abundance exhibit
overabundances of Sr from 0.5 to 1.35~dex, while the star with the lowest C
abundance exhibits an underabundance of Sr by 0.5 dex. Barium is overabundant
in all three stars. The Ba abundances were estimated from the Ba {\sc ii}
493.4 nm and 614.1 nm lines. For the analysis we adopted the hyperfine
splitting by McWilliam (1998) and assumed a solar isotopic mix. The measured Ba and Sr
abundances are listed in Table~\ref{tab1}. The analysis was performed using 1D LTE models.

\begin{figure}[ht]
\centering \includegraphics{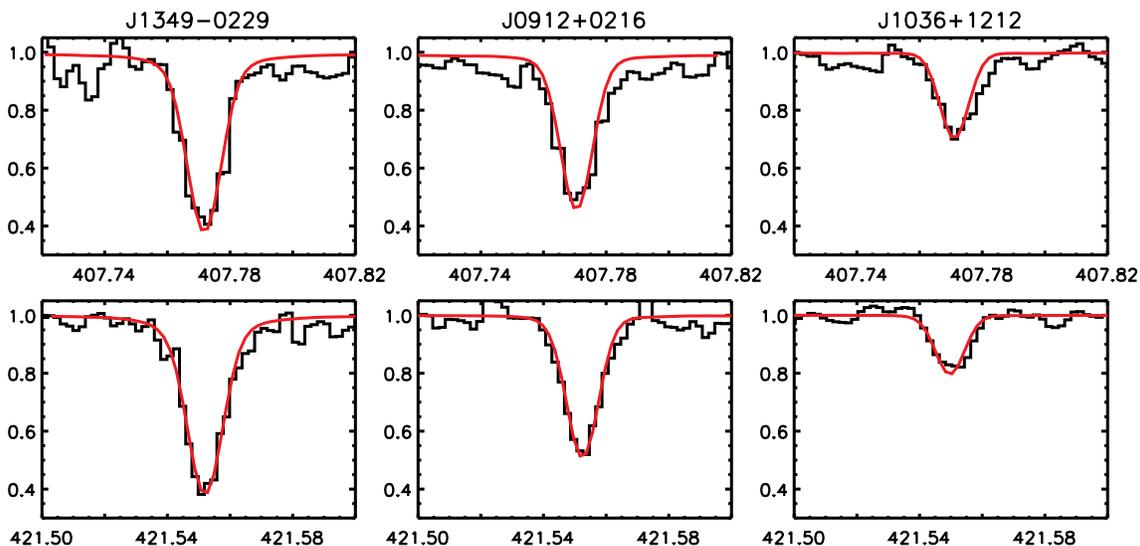}
\caption{Sr {\sc ii} 407.7 nm and 421.5 nm lines used in the
  abundance analysis.}
\label{fig3}
\end{figure}

\section{Discussion and conclusions}

The 3D effects on the carbon abundance determined
using CH lines were found to be significant, with 3D corrections of up to
--0.7 dex. This suggests that all the published
C abundances in dwarf CEMP stars, based on molecular lines
and 1D models should be reconsidered in the light of hydrodynamic simulations.
Star SDSS J1349-0229 exhibits a strong C$_2$ Swan band which
classifies it as a carbon star (C/O$>1$), while the 
other two stars are carbon-enhanced stars.

\begin{figure}
\includegraphics{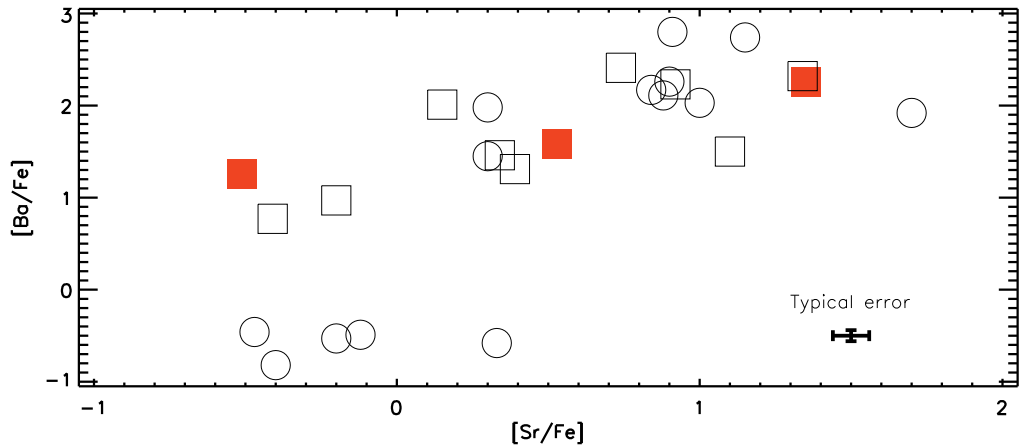}
\caption{[Ba/Fe] vs. [Sr/Fe] for our three CEMP stars (red filled
  symbols) plotted alongside values for 24 CEMP stars (open squares for
  dwarfs, circles for giants) listed in Table 4 of Sivarani et al.~(2006). } 
\label{fig4}
\end{figure}

In Figure~\ref{fig4} we plot the Sr and Ba abundances
of the three stars (red symbols) together with a sample of
24 CEMP stars from Table 4 of Sivarani et al.~(2006).
It is now widely accepted that the CEMP class includes
objects which have rather diverse astrophysical origin.
One sub-classification of such stars was proposed
by Beers \& Christlieb~(2005) based on the
abundances of neutron-capture elements in 
CEMP stars. One therefore speaks of
CEMP-no stars, when no enhancement of
neutron capture elements is observed;
CEMP-s when only the s-process elements
are enhanced, and CEMP-r/s when both
r-process and s-process elements are enhanced. 
The class CEMP-no/s was introduced 
by Sivarani et al.~(2006) to contain
the stars CS~29527-041 and CS~31080-095
which where overabundant in Ba but underabundant
in Sr. Star SDSS J1036+1212 is the third star 
known to fall into this category.
Interestingly all three known CEMP-no/s stars
are warm dwarf/TO stars. The vast majority
of CEMP dwarf stars is understood as the result
of mass-transfer from a presently unseen companion
which has undergone the AGB phase.
The peculiar Sr/Ba ratios of the CEMP-no/s stars
should provide useful constraints on the 
properties of the companion star while on the AGB.

The other two stars, SDSS J0912+0216 and SDSS J1349-0229,
exhibit both high abundances of
Sr and Ba, indicating that they are CEMP-s stars.
Our single epoch observations offer no indications
on the possible existence of binary companions, 
however radial velocity monitoring of all three
would be highly desirable.

\end{document}